\documentclass[showpacs,aps,twocolumn,graphicx]{revtex4}%
\usepackage{graphicx}

\def\<{\langle}
\def\>{\rangle}

\begin{document}

\title{Circular quantum secret sharing}

\author{ Fu-Guo Deng,$^{1,2,3,4}$\footnote{ E-mail addresses: fgdeng@bnu.edu.cn}  Hong-Yu Zhou,$ ^{1,2,3}$ and
Gui Lu Long$^{4,5}$ \footnote{ E-mail addresses:
gllong@tsinghua.edu.cn}}
\address{$^1$ The Key Laboratory of Beam Technology and Material
Modification of Ministry of Education, Beijing Normal University,
Beijing 100875,
China\\
$^2$ Institute of Low Energy
Nuclear Physics, and Department of Material Science and Engineering, Beijing Normal University, Beijing 100875,  China\\
$^3$ Beijing Radiation Center, Beijing 100875,  China\\
$^4$ Key Laboratory For Quantum Information and Measurements
of Ministry of Education, and Department of Physics, Tsinghua University, Beijing 100084, China\\
$^{5}$ Key Laboratory for Atomic and Molecular Nanosciences,
Tsinghua University, Beijing 100084, China }
\date{\today }

\begin{abstract}
A circular quantum secret sharing protocol is proposed, which is
useful and efficient when one of the parties of secret sharing is
remote to the others who are in adjacent, especially the parties
are more than three. We describe the process of this protocol and
discuss its security when the quantum information carrying is
polarized single photons running circularly. It will be shown that
entanglement is not necessary for quantum secret sharing.
Moreover, the theoretic efficiency is improved to approach 100\%
as almost all the instances can be used for generating the private
key, and each photon can carry one bit of information without
quantum storage. It is straightforwardly to utilize this
topological structure to complete quantum secret sharing with
multi-level two-particle entanglement in high capacity securely.
\end{abstract}
 \maketitle

\section{Introduction}

Secret sharing is a useful tool in classical secure communication
\cite{Blakley,HBB99}. It can be used to accomplish a special task.
Suppose a president of a bank, Alice wants to send a secret message
to her two agents, Bob and Charlie who are at a distant place for
carrying out a business on her behalf. Alice cannot determine
whether both of them are honest. There may be at most one dishonest
agent among Bob and Charlie pair, but Alice does not know who the
dishonest one is. She knows that the honest one will keep the
dishonest one from destroying the business if they coexist in the
process of the business. For the security of the secret message
($S_A$), she will split it into two pieces, $S_B$ and $S_C$, and
only sends one to Bob and another to Charlie respectively. The two
pieces of message can be used to reconstruct the secret message
$S_A$ when Bob and Charlie collaborate, otherwise none of them can
get any useful information about $S_A$.

Cryptography can be used to complete the task if Alice created a
private key $K_B$ ($K_C$) with Bob (Charlie). For example, if Alice
wants to send the message $S_B$ to Bob securely, she can encrypt the
message $S_B$ with the private key $K_B$ using the one time-pad
crypt-system and then send the ciphertext $C_B=S_B\oplus K_B$ to
Bob, where $\oplus$ means modulo 2 summation. With the key $K_B$,
Bob can decrypt $C_B$ to read out the message $S_B$, but any one
else can not obtain anything about it. Similarly, Alice can encrypt
the message $S_A$ with $K_A=K_B\oplus K_C$, i.e., $C_A=S_A\oplus
K_A$, and then sends the ciphered text $C_A$ to both of Bob and
Charlie. They can get the original message $S_A$ only when they
collaborate. In essence, this is the classical secret sharing whose
security depends on the privacy of the key $K_B$ and $K_C$. The
distribution of a private key between two remote parties or
multi-parties is important for secure communication.

Quantum key distribution (QKD) is  an important application of
quantum mechanics within the field of information, and it provides
a secure way for generating a private key between two remote
parties since Bennett and Brassard \cite{Gisin,BB84}. The secret
sharing has been generalized to the quantum scenario by using
entanglement \cite{HBB99,KKI}, namely quantum secret sharing
(QSS). Different from classical secret sharing, the shared
information in QSS can be both classical and quantum. In
particular, QSS is useful for creating a private key among
multi-party of secure communications . There have been many
theoretical and experimental interests in QSS
\cite{cleve,gottesman,Bandyopadhyay,aca,Karimipour,Tyc,guoqss,
Bagherinezhad,Sen,longqss,delay,Peng,MZ,dengmQSTS,dengpra,TZG,AMLance}.
A pioneering QSS scheme \cite{HBB99}, called HBB99 scheme, was
proposed by Hillery, Bu\v{z}ek and Berthiaume in 1999 by using
three-particle entangled Greenberger-Horne-Zeilinger (GHZ) states.
In this scheme, the bank president, Alice prepares a GHZ triplet
state
\begin{equation}
\vert \psi\rangle_{abc}=\frac{1}{\sqrt{2}}(\vert
000\rangle_{abc}+\vert 111\rangle_{abc}), \label{state1}
\end{equation}
where the state $\vert 0\rangle$ and $\vert 1\rangle$ are two
eigenvectors of two-level quantum system, such as the polarization
of single photons along the z-direction ($\sigma_z$). Alice sends
the particle $b$ and $c$ to Bob and Charlie respectively, and
keeps the particle $a$. They all agree that they choose randomly
one of the two measuring bases (MBs), $\sigma_x$ and $\sigma_y$ to
perform the measurement on their particles. When they all choose
$\sigma_x$ or one chooses $\sigma_x$ and the others choose
$\sigma_y$, their results are correlated and will be kept for key,
otherwise they discard the results. Its intrinsic efficiency for
qubits $\epsilon_q$, the ratio of number of valid qubits to the
number of transmitted qubits, is about 50\% as half of the
instances will be abandoned. Subsequently, Karlsson, Koashi and
Imoto (KKI) put forward a QSS scheme \cite{KKI} with two-photon
polarization-entangled states, and its efficiency $\epsilon_q$ is
also 50\%.

There is a common feature in the existing QSS protocols, for
instance, in Refs.
\cite{HBB99,KKI,cleve,gottesman,Bandyopadhyay,aca,Karimipour,Tyc,guoqss,
Bagherinezhad,Sen,longqss,delay,Peng,MZ,TZG}, that the quantum
information carrier (QIC) runs only between two parties among the
participants, i.e., from Alice to Bob, and from Alice to Charlie
in three-party secret sharing, no transmission between Bob and
Charlie. In general, Alice is remote to both Bob and Charlie, and
Bob and Charlie are likely two adjacent agents. Then they can
complete a QSS by making the QIC run a circle, namely, the QIC
runs from Alice to Bob, and then from Bob to Charlie, finally back
to Alice. Though the change in the process seems small, but it
reduces the resource requirement greatly and the intrinsic
efficiency is also increased. In this paper, we will present a
quantum secret sharing protocol of classical information  based on
the circular motion idea using polarized single photons. This
basic circular transmission idea is not restricted to the use of
single photons, but also could be used in other systems and the
generalization of the protocol with entangled states is also
presented.

\section{Circular quantum secret sharing with polarized single photons}
\subsection{The circular-QSS protocol with single photons}

The basic idea of the circular-QSS protocol with polarized single
photons is shown in Fig.1. The president, Alice prepares the QICs
which are polarized single photons in this QSS protocol, using two
sets of measuring basis (MB) into one of the following four states
randomly
\begin{eqnarray} \{\vert +z\rangle, \vert
-z\rangle, \vert +x\rangle, \vert -x\rangle\},\end{eqnarray} where
\begin{eqnarray}
\vert +z\rangle&=&\vert 0\rangle, \,\,\,\, \vert -z\rangle=\vert
1\rangle, \label{state2}\\
 \vert
+x\rangle&=&\frac{1}{\sqrt{2}}(\vert 0\rangle+\vert 1\rangle),
 \,\,\,\, \vert -x\rangle=\frac{1}{\sqrt{2}}(\vert 0\rangle-\vert
 1\rangle), \label{state3}
\end{eqnarray}
respectively. The states $\vert \pm z\rangle$ and $\vert \pm
x\rangle$ are the eigenstates of $\sigma_z$ and $\sigma_x$,
respectively. Before the quantum communication, Alice, Bob and
Charlie agree that Bob and Charlie choose randomly the control
mode or the coding mode for the quantum signal received, similar
to the Ping-Pong quantum key distribution scheme in Refs.
\cite{pingpong,cai2}. When they choose the control mode, they
perform single-photon measurement on the signal using one of the
two MBs, $\sigma_z$ and $\sigma_x$ randomly and record the MBs and
outcomes of the measurements, denoted as  $R_B$ and $R_C$
respectively. If they choose the coding mode, they perform
randomly one of the two unitary operations, $U_0$ and $U_1$ which
represent the bits 0 and 1 respectively, on the single photon
received,
\begin{eqnarray}
U_0&=&\vert 0\rangle\langle 0\vert+\vert 1\rangle\langle 1\vert,
\label{o1}\\
U_1&=&\vert 0\rangle\langle 1\vert-\vert 1\rangle\langle 0\vert.
\label{o2}
\end{eqnarray}

The operation $U_0$ is the identity operation and does nothing on
the single photon. The nice feature of the $U_1$ operation is that
it flips the state in both measuring basis:  $U_1$ negates the
states in the two conjugate MBs\cite{QOTP,BidQKD}, i.e.,
\begin{eqnarray}
U_1\vert +z\rangle&=&-\vert -z\rangle, \,\,\,\, U_1\vert
-z\rangle=\vert +z\rangle,\\
U_1\vert +x\rangle&=&\vert -x\rangle, \,\,\,\,  U_1\vert
-x\rangle=-\vert +x\rangle.
\end{eqnarray}
\begin{figure}[!h]
\begin{center}
\includegraphics[width=8cm,angle=0]{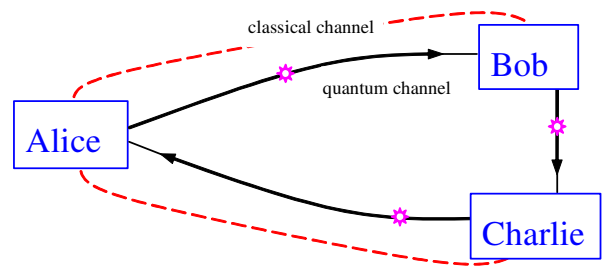} \label{f1}
\caption{ Circular quantum secret sharing. The quantum information
carrier runs from Alice to Bob, then to Charlie, and finally back
to Alice. Bob and Charlie randomly choose the control mode or the
coding mode for each signal. The full line represents the quantum
channel and the dashed line for classical channel.}
\end{center}
\end{figure}

For creating the private key $K_A$, Alice sends the quantum signal
to Bob first,  and Bob chooses the control mode or the coding mode
randomly for each photon. If he chooses the control mode, Bob
performs measurement on the photon. Otherwise, he codes the photon
with the two unitary operations $U_0$ and $U_1$ chosen randomly
and then sends it to Charlie. Charlie performs the operation in a
way just like Bob. He sends the photon to Alice after coding with
unitary operations if he chooses the coding mode. Alice performs
single-photon measurement on the photon received with the same MB
as she prepares it. As the two unitary operations $U_0$ and $U_1$
do not change the MB, Alice can get a deterministic outcome for
each photon returned, e.g., $U_A=U_B\otimes U_C$, where $U_B$ and
$U_C$ are the operations done by Bob and Charlie on the same
photon respectively, $U_A$ is the total operation on photon.

In order to prevent any eavesdropper from getting the information
about the key $K_A$ which is represented by $U_A$ when the parties
confirm that the whole process of quantum communication is secure,
Alice, Bob and Charlie should sample instances twice for analyzing
the error rates. The first sequence of the samples are those that
have been chosen and measured by  Bob or Charlie when they choose
the control mode. It can be divided into two parts: one contains
those measured by Bob, denoted by $s_{1b}$; the other contains
those measured  by Charlie, say $s_{1c}$. The second sample
sequence are randomly chosen from those instances that returned
back to Alice, denoted by $s_2$. In this second sample sequence,
Bob and Charlie both choose the coding mode, and hence accordingly
their coding operations form two coding sequences denoted by
$s_{2b}$ and $s_{2c}$.

For analyzing the error rate in samples $s_{1b}$,  Bob publishes the
information about the samples $s_{1b}$, including the positions of
the measured photons in the photon sequence he receives from Alice,
the MBs and the results of the measurements in $s_{1b}$, and Alice
compares it with the information of these photons. In those cases
where Bob has chosen the same measuring-basis as Alice's, Alice can
determine the error rate $\varepsilon_{_{1b}}$.

The error rate of the sample sequence $s_{1c}$, denoted as
$\varepsilon_{_{1c}}$, can be similarly determined by Alice and
Charlie. But here Charlie first announces the positions of
$s_{1c}$ photons, and then Bob is asked to publish his unitary
operations he performs on these photons, then Bob publishes the
outcome of his measurement and the corresponding measuring-basis.

To analyze the error rate in the second sample sequence $s_2$,
Alice asks Bob and Charlie to publish the unitary operations of
the sampled photons in the sequence. Since Alice's result should
be the product of unitary operations of Alice and Bob, Alice can
determine the error rate from the announcement of Bob and Charlie.

With these eavesdropping checks, eavesdropper will be detected if
he or she has monitored the quantum channel. The details will be
discussed shortly.

In practical, there are noise and loss in the quantum channel. The
methods for error correction and privacy amplification are
necessary for distilling the key $K_A=K_B \oplus K_C$, same as QKD
\cite{Gisin,book}.

\subsection{Security analysis of the QSS protocol with single photons}

Suppose the dishonest one between Bob and Charlie is denoted as
Bob*. As discussed in Ref. \cite{KKI}, if the dishonest one Bob*
can be detected by the other two parties, say Alice and Charlie*
when he/she eavesdrops the quantum communication, then any
eavesdropper can be found out. We will discuss the security of
this circular QSS protocol in two cases with Bob* being Bob, and
Charlie respectively.

If Bob* is Bob, the security of this QSS protocol is simplified to
prevent Bob from eavesdropping the secret key $K_A$. In fact, the
task of eavesdropping check is to determine whether Bob obtained
the information about the unitary operations $U_C$ which is just
the key $K_C$ under ideal condition. Any other cheat done by Bob
in the process of quantum communication will be found out in
secret sharing if Bob cannot get the information about the $U_C$.
The fake information that Bob publishes about his operations $U_B$
on $s_{1c}$ will be exposed as it was announced before Charlie
publishes the MBs and the results for $s_{1c}$. In this way, the
process for security analysis between Alice and Charlie is equal
to that in BB84 QKD \cite{BB84}, which is proved unconditionally
secure, for example, seeing Refs.
\cite{BB84security1,BB84security2,BB84security3}. We can calculate
the information $I_{B}$ that Bob can obtain about the unitary
operations $U_C$ done by Charlie with the probability of being
detected $\varepsilon_{_{B}}$ as follows, in a way similar to
those used in  Refs. \cite{pingpong,twostep}.

We discuss the security in the case that any eavesdropper can only
make individual attacks. The reason is discussed in Refs.
\cite{attack1,attack2,attack3,attack4}.  As discussed in Ref.
\cite{BidQKD}, the limitation on the error rate introduced by
Bob's eavesdropping is 25\% for which Bob intercepts the quantum
signal Alice sends to Charlie and sends a fake photon to him. The
purpose that Bob eavesdroppers the quantum signal is to learn more
information about it and introduces as little  error as possible
into the results. The error rate introduced by Bob comes from the
wrong MBs chosen, i.e., Alice prepares the quantum signal with
$\sigma_z$, but Bob chooses $\sigma_x$ for eavesdropping, or vice
versa \cite{attack1}. We assume that Alice prepares the quantum
states with $\sigma_z$ and Bob with $\sigma_x$ for eavesdropping
(the condition that Alice chooses $\sigma_x$ and Bob $\sigma_z$ is
the same for the security analysis). In this way, the information
stolen by Bob about the state of the photon coded is equal to that
about the operation done by Charlie \cite{pingpong}.

The optimal individual attack done by an eavesdropper can be
realized by a unitary operation $U_E$ on the photon
\cite{pingpong,book,attack1,attack2,attack3,attack4,Preskill,W1,D1,LM}
with an ancilla whose initial state is $\vert 0\rangle$.
\begin{eqnarray}
U_E\vert 0\rangle\vert 0\rangle&=&\vert 0\rangle\vert 0\rangle,\\
U_E\vert 1\rangle\vert 0\rangle&=&\cos\phi\vert 1\rangle\vert
0\rangle+\sin\phi\vert 0\rangle\vert 1\rangle,
\end{eqnarray}
where $\phi\in[0,\frac{\pi}{4}]$ characterizes the strength of
Eve's attack \cite{attack2}.

The probability $\varepsilon_{_{B}}$ that Bob will be detected is
same as the error rate introduced by the
eavesdropping\cite{pingpong,book}. As Alice makes the photon in
the four states $\{\vert +z\rangle, \vert -z\rangle, \vert
+x\rangle, \vert -x\rangle\}$ with the same probability, then the
error rate is \cite{BidQKD,attack2}
\begin{equation}
\varepsilon_{_{B}}=\frac{1}{2}\sin^2\phi.
\end{equation}

The state of the photon that Alice prepares can be described with
a density matrix
\begin{equation}
\rho_A=\frac{1}{2}\vert 0\rangle\langle 0\vert+\frac{1}{2}\vert
1\rangle\langle 1\vert.
\end{equation}
After Bob's eavesdropping, the joint state $\psi_{\Im}$ of the
system $\Im$ composed of the photon $A$ and the ancilla
\cite{pingpong,book,twostep} $P$ can be written as
\begin{eqnarray}
\rho_{_{AP}}&=&\frac{1}{2}\{\vert 00\rangle\langle 00\vert +
\cos^2\phi\vert 10\rangle\langle 10\vert + cos\phi sin^*\phi \vert
10\rangle\langle 01\vert \nonumber\\ &+& sin\phi cos^*\phi \vert
01\rangle\langle 10\vert + sin^2\phi \vert 01\rangle\langle
01\vert\},
\end{eqnarray}
where $\vert ij\rangle \equiv \vert i\rangle_A\vert j\rangle_P$,
$i,j\in \{0,1\}$. The effect of the unitary operations done by
Charlie is just to change the state of the photon $A$ in
$\rho_{_{AP}}$. Suppose the probabilities that Charlie chooses
$U_0$ and $U_1$ are $P_{c0}$ and $P_{c1}$, respectively. After the
coding, the state $\psi_{\Im}$ becomes
\begin{eqnarray}
\rho'_{_{AP}}&=&\frac{1}{2}\{P_{c0}\vert 00\rangle\langle 00\vert
+ P_{c1}\vert 10\rangle\langle 10\vert \nonumber\\ &+&
P_{c0}[\cos^2\phi\vert 10\rangle\langle 10\vert + \cos\phi
\sin^*\phi \vert 10\rangle\langle 01\vert \nonumber\\ &+& \sin\phi
\cos^*\phi \vert 01\rangle\langle 10\vert  + \sin^2\phi \vert
01\rangle\langle 01\vert] \nonumber\\ &+& P_{c1}[\cos^2\phi\vert
00\rangle\langle
00\vert - \cos\phi sin^*\phi \vert 00\rangle\langle 11\vert \nonumber\\
 &-& \sin\phi \cos^*\phi \vert
11\rangle\langle 00\vert + \sin^2\phi \vert 11\rangle\langle
11\vert]\}.
\end{eqnarray}
As Bob wants to eavesdrop the quantum communication for creating a
private key, he should send the photon coded to Alice and only
measure the ancilla, which is different to that for direct
communication \cite{pingpong,twostep}. We can trace out the state
of the photon $A$ from the joint state $\psi_{\Im}$ with MB
$\sigma_z$ ($\{ \vert 0\rangle, \vert 1\rangle \}$) to get the
state of the ancilla, $\rho'_{_{P}}$,
\begin{eqnarray}
\rho'_{_{P}}=\frac{1}{2}\{(1+cos^2\phi)\vert 0\rangle\langle
0\vert +sin^2\phi\vert 1\rangle\langle 1\vert\},
\end{eqnarray}
which can be projected to orthogonal measuring basis $\{\vert
0\rangle,\vert 1\rangle\}$ (it is one of the best measurements for
distilling the information from the state) and written as
\begin{eqnarray}
\rho''_{_{P}}=\frac{1}{2}\left(
\begin{array}{cc}
1 + cos^2\phi & 0 \\
0 & sin^2\phi%
\end{array}%
\right).
\end{eqnarray}

\begin{figure}[!h]
\begin{center}
\includegraphics[width=8cm,angle=0]{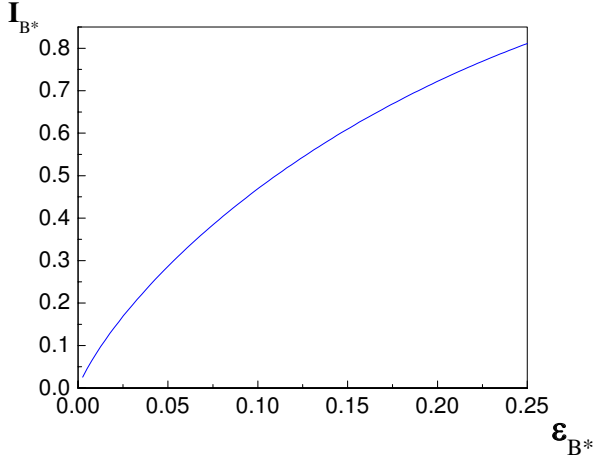} \label{fig2}
\caption{ The relation between $I_{_{B}}$ and
$\varepsilon_{_{B}}$.}
\end{center}
\end{figure}

The information $I_{B}$ that Bob can obtain is equal to the Von
Neumann entropy of the state of the ancilla. And the Von Neumann
entropy can be calculate as follows \cite{book,BidQKD}.
\begin{eqnarray}
I_{_{B}}=S(\rho''_{_{P}})=-Tr(\rho''_{_{P}}log_2\rho''_{_{P}}),
\end{eqnarray}
i.e.,
\begin{eqnarray}
I_{_{B}}=S(\rho''_{_{P}})=-\sum_{i=0}^1\lambda_ilog_2\lambda_i,
\end{eqnarray}
where $\lambda_i$ ($i=0,1$) are the roots of the characteristic
polynomial $det(\rho''_{_{P}}-\lambda I)$ \cite{pingpong},
yielding the two eigenvalues
\begin{eqnarray}
\lambda_0=\frac{1}{2}(1+\cos^2\phi),
\end{eqnarray}
\begin{eqnarray}
\lambda_1=\frac{1}{2}\sin^2\phi.
\end{eqnarray}
So we have
\begin{eqnarray}
I_{_{B}}&=&1-\frac{1}{2}\{(1+\cos^2\phi)\log_2(1+\cos^2\phi) +
\sin^2\phi \log_2 \sin^2\phi\} \nonumber\\
&=&-\varepsilon_{_{B}}log_2\varepsilon_{_{B}}-(1-\varepsilon_{_{B}})\log_2(1-\varepsilon_{_{B}}).
\end{eqnarray}
 The relation between
$I_{_{B}}$ and $\varepsilon_{_{B}}$ is shown in Fig.2. It is shown
in the figure that Bob has to face a detection probability
$\varepsilon_{_{B}}>0$ if he wants to gain information
$I_{_{B}}>0$. If $I_{_{B}}$ is not small, Bob will be detected,
otherwise Alice and Charlie can distill the key $K_C$ with privacy
amplification.

If the Bob* is Charlie, the process of eavesdropping check for the
sample sequence $s_{1b}$ is same as in the BB84-QKD protocol
\cite{BB84,BB84security1,BB84security2,BB84security3}. The state
$\rho_{A''}$ of the photon prepared by Alice is random for Charlie
as she chooses randomly one of the two MBs $\sigma_z$ and
$\sigma_x$ for it.
\begin{eqnarray}
\rho_{_{A''}}&=&\frac{1}{4}\vert +z\rangle\langle +z\vert +
\frac{1}{4}\vert -z\rangle\langle -z\vert + \frac{1}{4}\vert
+x\rangle\langle +x\vert \nonumber\\
 &+& \frac{1}{4}\vert -x\rangle\langle -x\vert
 = \frac{1}{2}\left(
\begin{array}{cc}
1 & 0 \\
0 & 1
\end{array}
\right).
\end{eqnarray}
The operation $U_B$ done by Bob on the photon does not change the
character of the state for Charlie as
\begin{eqnarray}
\rho_{_{A'''}}&=&\frac{P_{b0}}{4}\vert +z\rangle\langle +z\vert +
\frac{P_{b1}}{4}\vert -z\rangle\langle -z\vert \nonumber\\ &+&
\frac{P_{b0}}{4}\vert -z\rangle\langle -z\vert +
\frac{P_{b1}}{4}\vert +z\rangle\langle +z\vert \nonumber\\ &+&
\frac{P_{b0}}{4}\vert +x\rangle\langle +x\vert +
\frac{P_{b1}}{4}\vert -x\rangle\langle -x\vert \nonumber\\ &+&
\frac{P_{b0}}{4}\vert -x\rangle\langle -x\vert +
\frac{P_{b1}}{4}\vert +x\rangle\langle
+x\vert\nonumber\\
 &=& \frac{1}{2}\left(
\begin{array}{cc}
1 & 0 \\
0 & 1
\end{array}
\right)=\rho_{_{A''}},
\end{eqnarray}
where $P_{b0}$ and $P_{b1}$ are the probabilities that Bob chooses
the unitary operations $U_0$ and $U_1$, respectively. No matter
what quantum signal Charlie eavesdrops, the security analysis is
same as BB84-QKD \cite{BB84}. So this QSS is secure if Bob* is
Charlie.

Bob* may  cheat in the communication, for instance he publishes a
wrong information about his unitary operations $U_{B*}$, or he
does not use the right key $K_{B*}$ in secret sharing. Inevitably,
his action can be detected. For example, the wrong information
about his unitary operations will be found out when Alice and
Charlie compare the results in $s_{2c*}$ in quantum secret
sharing. After the key $K_A=K_B\oplus K_C$ is created, the cheat
that Bob* does not use $K_{B*}$ for decrypting the ciphered text
$C_A=S_A\oplus K_A$ can also be detected before the $C_A$ is
transmitted. Alice, Bob and Charlie need only determine whether
the key $K'_A=K'_B\oplus K'_C$ obtained by combining Bob's key and
Charlie's key when they cooperate is identical to her key $K_A$
obtained by Alice's measurement before she sends a secret message
to her two remote assistants, Bob and Charlie. The process can be
achieved by choosing at random a sufficiently large subset of bits
in the key $K'_A$ to compare the results with those in the key
$K_A$. If the error rate is zero, Alice confirms that there is no
dishonest one in Bob and Charlie pair, and she sends the secret
message to them after encrypting it with the key $K_A$; otherwise
she has to abort the secret message communication.

The parties encode the photons with unitary operations and each
photon can carry one bit of information in $K_A$ in principle. The
efficiency for qubit is improved to approach 100\%. Moreover, they
do not exchange the information about the MBs for almost all the
instances, and they also need not store the single photons.

\section{Circular Quantum Secret Sharing with multi-level two-particle entanglement}

For two-particle quantum system, $d$-dimension Bell-basis states
in a symmetric quantum channel are
\cite{BW,bennett,longzeng,superdense1,superdense2}
\begin{equation}
\vert \Psi_{nm}\rangle =\sum_{j} e^{2\pi ijn/d}\vert j\rangle
\otimes \vert j+m\;{\rm mod} \; d \rangle/\sqrt{d},
\end{equation}
where $n,m=0,1,...,d-1$. The unitary operations
\begin{equation}
U_{nm} =\sum_{j} e^{2\pi ijn/d}\vert j+m\;{\rm mod} \; d \rangle
\langle j\vert
\end{equation}
can transform the Bell-basis state
\begin{equation}
\vert \Psi_{00}\rangle =\sum_{j} \vert j\rangle \otimes \vert j
\rangle/\sqrt{d}
\end{equation}
into the Bell-basis state $\vert \Psi_{nm}\rangle$, i.e.,
$U_{nm}\vert \Psi_{00}\rangle=\vert \Psi_{nm}\rangle$. For
two-party communication, one particle can carry $\log_{2}d^{2}$
bits of information while running forth and back. In a more
generalized case,  non-symmetric quantum channel is possible where
the two particles of the entangled quantum system have the
different dimensions \cite{superdense2,yan}, for example, the
first particle has $p$ dimensions and the second one has $q$
dimensions. Then the capacity is $\log_{2}pq$.

The source coding capacity of this circular QSS can be improved
largely with super-dense coding \cite{BW,superdense1,superdense2}
and quantum state storage \cite{storage1,storage2,sun}. We will
generalize this circular QSS with Einstein-Podolsky-Rosen (EPR)
pairs, two-particle maximally entangled states, following the
ideas in dense coding \cite{BW}. The case for other multi-level
two-particle entanglement is just the same as it.

An EPR pair can be in one of the four Bell states
\cite{book,longliu},
\begin{eqnarray}
\left\vert \psi ^{-}\right\rangle_{HT}
=\frac{1}{\sqrt{2}}(\left\vert 0\right\rangle_{H}\left\vert
1\right\rangle_{T} - \left\vert 1\right\rangle_{H}\left\vert
0\right\rangle_{T}), \label{EPR1}\\
\left\vert \psi ^{+}\right\rangle_{HT}
=\frac{1}{\sqrt{2}}(\left\vert 0\right\rangle_{H}\left\vert
1\right\rangle_{T} + \left\vert 1\right\rangle_{H}\left\vert
0\right\rangle_{T}), \label{EPR2}\\
\left\vert \phi ^{-}\right\rangle_{HT}
=\frac{1}{\sqrt{2}}(\left\vert 0\right\rangle_{H}\left\vert
0\right\rangle_{T} - \left\vert 1\right\rangle_{H}\left\vert
1\right\rangle_{T}), \label{EPR3}\\
\left\vert \phi ^{+}\right\rangle_{HT}
=\frac{1}{\sqrt{2}}(\left\vert 0\right\rangle_{H}\left\vert
0\right\rangle_{T}+\left\vert 1\right\rangle_{H}\left\vert
1\right\rangle_{T}).  \label{EPR4}
\end{eqnarray}
The four local unitary operations $U_{Li}$ ($i=0,1,2,3$) can
transfer the four Bell states into each other.
\begin{eqnarray}
U _{L0}&=&I=\left\vert 0\right\rangle \left\langle 0\right\vert
+\left\vert 1\right\rangle \left\langle 1\right\vert,
\label{O0}\\
U _{L1}&=&i\sigma _{y}=\left\vert 0\right\rangle \left\langle
1\right\vert -\left\vert 1\right\rangle \left\langle 0\right\vert,
\label{O2}\\
U _{L2}&=&\sigma _{x}=\left\vert 1\right\rangle \left\langle
0\right\vert +\left\vert 0\right\rangle \left\langle 1\right\vert,
\label{O1}\\
U _{L3}&=&\sigma _{z}=\left\vert 0\right\rangle \left\langle
0\right\vert -\left\vert 1\right\rangle \left\langle 1\right\vert,
\label{O3}
\end{eqnarray}
i.e.,
\begin{eqnarray}
&&I \otimes U _{L0}\vert \psi^\pm\rangle=\vert \psi^\pm\rangle,
\,\,\,\, I \otimes U _{L0}\vert \phi^\pm\rangle=\vert
\phi^\pm\rangle,\label{L0}\\
&&I \otimes U _{L1}\vert \psi^\pm\rangle=\vert \phi^\mp\rangle,
\,\,\,\, I \otimes U _{L1}\vert \phi^\pm\rangle=-\vert
\psi^\mp\rangle,\label{L1}\\
&&I \otimes U _{L2}\vert \psi^\pm\rangle=\vert \phi^\pm\rangle,
\,\,\,\, I \otimes U _{L2}\vert \phi^\pm\rangle=\vert
\psi^\pm\rangle,\label{L2}\\
&&I \otimes U _{L3}\vert \psi^\pm\rangle=-\vert \psi^\mp\rangle,
\,\,\,\, I \otimes U _{L3}\vert \phi^\pm\rangle=\vert
\phi^\mp\rangle.\label{L3}
\end{eqnarray}
The process of this QSS with EPR pairs is similar to that with
single photons discussed above. The president, Alice prepares the
two-particle entangled state $\vert \psi^-\rangle_{HT}$, and she
keeps the particle $H$ and sends the particle $T$ to Bob first,
shown in Fig.1. He chooses the control mode and the coding mode
randomly. When Bob chooses the coding mode, he performs one of the
four unitary operations $U_{Li}$ ($i=0,1,2,3$) which represent the
bits 00, 01, 10 and 11 respectively, on the particle $T$ randomly.
Otherwise, he chooses the two MBs, $\sigma_z$ and $\sigma_x$ to
measure the particle $T$, and tells Alice which particle he
chooses the control mode. Alice does the correlated measurement on
the particle $H$ in the EPR pair in which Bob measures the
particle $T$, that is, Bob tells Alice the position of the
particle and his MB for it, and Alice performs the measurement
with the same MB as Bob on the particle $H$.

As for Charlie,  after he receives the particle Bob sends  to
Charlie after coding with an unitary operation, Charlie chooses
randomly the control mode and the coding mode. If he chooses the
coding mode, he performs randomly one of the four coding
operations and then send the particle to Alice. When Bob chooses
the control mode, he measures the particle choosing randomly one
of the two MBs $\sigma_z$ and $\sigma_x$.

The eavesdropping check can be adapted here straightforwardly. In
fact, no matter who the dishonest Bob* is, the way for checking
the security of quantum communication is same as that for the
BBM-QKD \cite{BBM92} which has been proven unconditionally secure
for key generation \cite{attack3,BBMsecurity2}. As pointed out by
Bechmann-Pasquinucci and Peres \cite{mulitilevel}, the QKD with
multi-level quantum system is more secure than that with two-level
one.

\section{Discussion and conclusion}

In general, QSS is accomplished with entanglement, which normally
requires more complicated experimental setups. Though big progress
has been made for producing and measuring entanglement, the
efficiency is still low
\cite{threeentanglement,fourentanglement,fiveentanglement}. QSS
with single photons will be more convenient for being implemented
in laboratory and practical application. On the other hand, the
source coding capacity of QSS can be improved largely with
super-dense coding in which entanglement is necessary. With
development of technology, it is likely feasible to implement QSS
based on entanglement, especially with multi-level entanglement in
high capacity.

Certainly, another important function of QSS is to split a secret
message into $n$ pieces and completes the task of an $m-out-of-n$
quantum secret splitting scheme, or so-called ($m$, $n$) threshold
scheme \cite{KKI}. Unfortunately, this circular QSS scheme can not
be used to accomplish the full goal of quantum secret splitting.
That is, it can not be used for $m-out-of-n$ scheme in which any
$m$ parties can reconstruct the secret message when they
collaborate. However, it is useful for accomplish a partial goal,
$n-out-of-n$ scheme. In other words, the circular QSS can be used
to reconstruct the secret message when all of the other $n$
parties cooperate with some classical information published by
Alice, the president.

In summary, a circular QSS scheme is proposed. It is useful and
efficient when the president Alice is remote to all her agents,
Bobs who are in adjacent, especially the parties of secret sharing
are more than three. In this scheme, the quantum information
carrier, single photons or entangled particles, will run
circularly, and the parties choose randomly the control mode or
coding mode to operate the QIC. They measure the QIC only when
they choose control mode, otherwise, they encode the QIC with some
unitary operations. If the QIC is single photon, all the parties
of communication including Alice do not need to store the quantum
state. If the QIC is entangled quantum system, only Alice is
required to possess the technique of quantum storage, others need
not. This is convenient for realizing QSS in practical
application. Moreover, each QIC can be used to carry information
except for the samples for eavesdropping check, and classical
information exchanged is reduced largely as the parties need not
announce the MBs for the QIC.

\bigskip
\section*{Acknowledgments}
This work is supported by the National Natural Science Foundation of
China under Grant Nos. 10604008, 10435020, 10254002, A0325401,
60433050 and 10325521, the National Fundamental Research Program
under Grant No. 001CB309308, the SRFDP program of Education Ministry
of China.

\end{document}